\documentclass[10pt,twocolumn]{article}
\usepackage{graphicx}
\usepackage{amssymb,amsthm,amsmath}
\usepackage{xcolor,paralist,titlesec,fancyhdr,etoolbox}
\usepackage[citecolor=black]{hyperref}
\usepackage[british]{babel}
\usepackage{natbib}
\let\cite\citep
\usepackage{caption}

\hypersetup{colorlinks=true, linkcolor=black, filecolor=black, urlcolor=black }

\usepackage[acronym]{glossaries-extra} %
\usepackage{booktabs} %
\usepackage{datatool}
\usepackage{siunitx}
\DTLloaddb[noheader, keys={thekey,thevalue}]{cleaned}{data/stats_post_QC.dat}
\DTLloaddb[noheader, keys={thekey,thevalue}]{filtered}{data/stats_RQ.dat}
\DTLloaddb[noheader, keys={thekey,thevalue}]{general}{data/general.dat}
\DTLloaddb[noheader, keys={thekey,thevalue}]{jsd}{data/jsd_stats.dat}
\DTLloaddb[noheader, keys={thekey,thevalue}]{walmm}{data/wa_lmm.dat}
\DTLloaddb[noheader, keys={thekey,thevalue}]{iglmm}{data/ig_lmm.dat}
\newcommand{\datclean}[1]{%
  \DTLgetvalueforkey{\dattmp}{thevalue}{cleaned}{thekey}{#1}%
  \num{\dattmp}%
}\newcommand{\datfiltered}[1]{%
  \DTLgetvalueforkey{\dattmp}{thevalue}{filtered}{thekey}{#1}%
  \num{\dattmp}%
}\newcommand{\generalvar}[1]{%
  \DTLgetvalueforkey{\dattmp}{thevalue}{general}{thekey}{#1}%
  \num{\dattmp}%
}\newcommand{\jsdvar}[1]{%
  \DTLgetvalueforkey{\dattmp}{thevalue}{jsd}{thekey}{#1}%
  \num{\dattmp}%
}\newcommand{\walmm}[1]{%
  \DTLgetvalueforkey{\dattmp}{thevalue}{walmm}{thekey}{#1}%
  \num{\dattmp}%
}\newcommand{\iglmm}[1]{%
  \DTLgetvalueforkey{\dattmp}{thevalue}{iglmm}{thekey}{#1}%
  \num{\dattmp}%
}

\usepackage{subcaption}

\usepackage{newfloat}
\usepackage{listings}

\begin{document}
\begin{filecontents*}{data/stats_post_QC.dat}
n_donations,195
n_chats,7047
n_messages,6527443
n_donations_wa,129
n_chats_wa,762
n_donations_ig,66
n_chats_ig,6285
mean_chats_donor,36.14
median_chats_donor,6
sd_chats_donor,103.07
min_chats_donor,5
max_chats_donor,1089
mean_chats_donor_wa,5.91
median_chats_donor_wa,5
sd_chats_donor_wa,2.38
min_chats_donor_wa,5
max_chats_donor_wa,21
mean_chats_donor_ig,95.23
median_chats_donor_ig,43
sd_chats_donor_ig,162.29
min_chats_donor_ig,6
max_chats_donor_ig,1089
mean_timespan_donor,1308
median_timespan_donor,1047
sd_timespan_donor,1048
min_timespan_donor,15
max_timespan_donor,4121
mean_timespan_donor_wa,1796
median_timespan_donor_wa,1907
sd_timespan_donor_wa,977
min_timespan_donor_wa,15
max_timespan_donor_wa,4121
mean_timespan_donor_ig,353
median_timespan_donor_ig,364
sd_timespan_donor_ig,35
min_timespan_donor_ig,181
max_timespan_donor_ig,365
mean_messages_donor,33474.07
median_messages_donor,14512
sd_messages_donor,49120.72
min_messages_donor,17
max_messages_donor,445036
mean_messages_donor_wa,47067.66
median_messages_donor_wa,29710
sd_messages_donor_wa,55324.08
min_messages_donor_wa,532
max_messages_donor_wa,445036
mean_messages_donor_ig,6904.77
median_messages_donor_ig,4331
sd_messages_donor_ig,9516.70
min_messages_donor_ig,17
max_messages_donor_ig,48508
\end{filecontents*}

\begin{filecontents*}{data/stats_RQ.dat}
n_donations,97
n_chats,889
n_messages,3423575
n_donations_wa,64
n_chats_wa,344
n_donations_ig,33
n_chats_ig,545
mean_chats_donor,9.16
median_chats_donor,5
sd_chats_donor,11.70
min_chats_donor,5
max_chats_donor,72
mean_chats_donor_wa,5.38
median_chats_donor_wa,5
sd_chats_donor_wa,0.68
min_chats_donor_wa,5
max_chats_donor_wa,7
mean_chats_donor_ig,16.52
median_chats_donor_ig,9
sd_chats_donor_ig,18.04
min_chats_donor_ig,5
max_chats_donor_ig,72
mean_timespan_donor,1197
median_timespan_donor,717
sd_timespan_donor,965
min_timespan_donor,38
max_timespan_donor,3814
mean_timespan_donor_wa,1630
median_timespan_donor_wa,1714
sd_timespan_donor_wa,927
min_timespan_donor_wa,38
max_timespan_donor_wa,3814
mean_timespan_donor_ig,357
median_timespan_donor_ig,364
sd_timespan_donor_ig,31
min_timespan_donor_ig,181
max_timespan_donor_ig,365
mean_messages_donor,35294.59
median_messages_donor,15751
sd_messages_donor,44088.91
min_messages_donor,871
max_messages_donor,187995
mean_messages_donor_wa,49006.33
median_messages_donor_wa,30213
sd_messages_donor_wa,48564.63
min_messages_donor_wa,1616
max_messages_donor_wa,187995
mean_messages_donor_ig,8702.12
median_messages_donor_ig,4617
sd_messages_donor_ig,9103.31
min_messages_donor_ig,871
max_messages_donor_ig,39882
\end{filecontents*}

\begin{filecontents*}{data/general.dat}
n_donations,97
n_chats,889
n_donations_wa,64
n_chats_wa,344
n_donations_ig,33
n_chats_ig,545
mean_chats_donor,9.16
median_chats_donor,5
sd_chats_donor,11.70
min_chats_donor,5
max_chats_donor,72
mean_chats_donor_wa,5.38
median_chats_donor_wa,5
sd_chats_donor_wa,0.68
min_chats_donor_wa,5
max_chats_donor_wa,7
mean_chats_donor_ig,16.52
median_chats_donor_ig,9
sd_chats_donor_ig,18.04
min_chats_donor_ig,5
max_chats_donor_ig,72
mean_timespan_donor,1197
median_timespan_donor,717
sd_timespan_donor,965
min_timespan_donor,38
max_timespan_donor,3814
mean_timespan_donor_wa,1630
median_timespan_donor_wa,1714
sd_timespan_donor_wa,927
min_timespan_donor_wa,38
max_timespan_donor_wa,3814
mean_timespan_donor_ig,357
median_timespan_donor_ig,364
sd_timespan_donor_ig,31
min_timespan_donor_ig,181
max_timespan_donor_ig,365
mean_messages_donor,35294.62
median_messages_donor,15750
sd_messages_donor,44088.88
min_messages_donor,871
max_messages_donor,187995
mean_messages_donor_wa,49006.38
median_messages_donor_wa,30215
sd_messages_donor_wa,48564.59
min_messages_donor_wa,1616
max_messages_donor_wa,187995
mean_messages_donor_ig,8702.12
median_messages_donor_ig,4617
sd_messages_donor_ig,9103.31
min_messages_donor_ig,871
max_messages_donor_ig,39882
\end{filecontents*}

\begin{filecontents*}{data/jsd_stats.dat}
mean,0.80
mean_wa,0.88
mean_ig,0.76
median,0.82
median_wa,0.90
median_ig,0.77
sd,0.11
sd_wa,0.07
sd_ig,0.10
skew,-0.83
skew_wa,-1.44
skew_ig,-0.69
kurtosis,0.42
kurtosis_wa,2.69
kurtosis_ig,0.39
\end{filecontents*}

\begin{filecontents*}{data/wa_lmm.dat}
intercept_est,0.115
intercept_ci_lo,0.083
intercept_ci_hi,0.146
intercept_se,0.016
intercept_df,285.439
intercept_t,7.083
intercept_p,0.000
rt_prev_alter_est,0.786
rt_prev_alter_ci_lo,0.739
rt_prev_alter_ci_hi,0.834
rt_prev_alter_se,0.024
rt_prev_alter_df,340.108
rt_prev_alter_t,32.498
rt_prev_alter_p,0.000
\end{filecontents*}

\begin{filecontents*}{data/ig_lmm.dat}
intercept_est,0.061
intercept_ci_lo,0.029
intercept_ci_hi,0.093
intercept_se,0.016
intercept_df,48.807
intercept_t,3.735
intercept_p,0.000
rt_prev_alter_est,0.796
rt_prev_alter_ci_lo,0.758
rt_prev_alter_ci_hi,0.834
rt_prev_alter_se,0.020
rt_prev_alter_df,541.629
rt_prev_alter_t,40.636
rt_prev_alter_p,0.000
\end{filecontents*}

\title{Sorry for the late reply: Response times and reciprocity in WhatsApp and Instagram chats.}
\author{Martin, Florian\footnote{Corresponding author: florian.martin@uni-bielefeld.de} \hspace{1cm} Hakobyan, Olya \hspace{1cm} Drimalla, Hanna\\
\small Center for Cognitive Interaction Technology (CITEC), Bielefeld University, Germany}
\maketitle

\begin{abstract}
Chat communication is often fast-paced, creating the expectation of quick replies. While the timing of exchanges is known to foster closeness and enjoyment, it remains largely unexplored whether chat partners with strong ties reciprocate each other's response times.
Using 3.4 million messages from \generalvar{n_chats} chats across \generalvar{n_donations} donations of anonymous WhatsApp and Instagram chats, we analyzed response times, their balance between chat partners, and its stability over time.
To our knowledge, this is the first study to examine response speed as an expression of reciprocity, bridging a key aspect of online communication with a fundamental principle of social interactions.
We found that around $70\%$ of WhatsApp and $44\%$ of Instagram messages were answered within five minutes, confirming the fast pace of instant messaging. Overall, the response speed between chat partners was similar. The response speed similarity was evident both in the overall response-time distributions of chat partners assessed with Jensen-Shannon distance and in the steep regression slopes (\walmm{rt_prev_alter_est} for WhatsApp and \iglmm{rt_prev_alter_est} for Instagram) linking one person's probability of responding within five minutes to the partner's corresponding probability.
Importantly, the dispersion of response time similarity over months showed that this balance persists over time. Our
results position response time balance as a marker of reciprocity in computer-mediated communication, offering a new way to quantitatively study this fundamental principle of social interaction.  We suggest using response speed balance as a complimentary metric in the analysis of relationship dynamics, such as the strengthening or weakening of social ties.

\end{abstract}

\setabbreviationstyle[acronym]{long-short}
\newcommand*{\glsfirstpl}[1]{%
  \glsentryplural{#1}%
  \glsadd{#1}%
}

\newacronym{GLMM}{GLMM}{generalized linear mixed model}
\newacronym{IM}{IM}{instant messaging}
\newacronym{JSD}{JSD}{Jensen-Shannon distance}
\newacronym{LMM}{LMM}{linear mixed-effects model}
\newacronym{RT}{RT}{response time}
\newacronym{RP}{RP}{response probability}
\newacronym{UUID}{UUID}{universally unique identifier}
\newacronym{SD}{SD}{standard deviation}
\newacronym{MAD}{MAD}{median absolute difference}
\newacronym{SMS}{SMS}{short messaging service}
\newacronym[\glslongpluralkey={percentage points}, \glsshortpluralkey={pp}]{pp}{pp}{percentage point}

\section{Introduction}

For most of human history, written correspondence meant waiting for a reply, often for weeks or even months. In contrast, modern \gls{IM} platforms, such as WhatsApp, WeChat and Telegram, are built on the promise of immediacy, allowing an exchange of text and media messages in  an almost conversational style. This shift in communication speed has implications for how response times are interpreted and what social signals they send. Prior work suggests that fast responses are common \cite{rosenfeld2018} and expected in chat communication \cite{pielot2014}, with slower responses being associated with negative interpersonal dynamics \cite{heston2017}.

In a highly connected world characterized by near-continuous availability, reply speed can be seen as a sign of how individuals allocate time and availability in social exchange. The expectations of responsiveness to chat messages point to a crucial aspect of human interaction: reciprocity. Reciprocity describes people's perceived inclination to return benefits that they have received. The study of reciprocity as a fundamental principle of human interaction predates chat communication by decades of theoretical and empirical work.  It has been called a "moral norm" for social interactions \cite{gouldner1960} and has been proposed to be a key aspect for relationships in theories of social exchange \cite{ahmad2023} and interpersonal adaptation \cite{burgoon2008}. Reciprocity has long been central to theories of social interaction, and its expression  in newer communication modes, such as instant messaging, invites further empirical investigation.

In recent years, the availability of digital communication data has enabled computational social science to revisit and expand classical theories of human interaction. One aspect of these efforts has been to examine whether known patterns of human interaction are reflected in computer-mediated communication. For instance, it has been shown that many properties of real-life interactions, such as social network size \cite{dunbar2015} and structure \cite{maccarron2016} or temporal properties, such as burstiness \cite{hakobyan2025}, are preserved in digital communication. For reciprocity specifically, it has been shown that people tend to reciprocate actions on Facebook in terms of "Like" reactions \cite{surma2016}. In romantic relationships, perceived texting similarity has been associated with greater relationship satisfaction \cite{ohadi2018}. On the more quantitative side, call records, text messages and social media logs have been used to quantify contact initiation or interaction volume between people, finding generally reciprocal interactions but also a substantial number of non-reciprocal ones \cite{kovanen2010,chowdhary2023,wang2013}. Beyond reflecting fundamental aspects of human interaction, such as reciprocity, digital trace data open new avenues by making fine-grained temporal aspects of social interactions observable at scale \cite{lazer2009}. Importantly, digital interaction traces can be collected through data donation \cite{boeschoten2022}, a participant-centered, privacy-preserving approach, in which individuals can voluntarily and anonymously share their messaging data for research purposes.

Building on methodological advances in computational social science and data donation, the present study uses messaging data traces to incorporate a temporal dimension into the analysis of reciprocity. Specifically, existing quantitative approaches to reciprocity have so far not taken into account the central aspect of currently popular IM services: pace of communication. In fact, over half of all messages on WhatsApp are answered as quickly as within one minute \cite{seufert2015, rosenfeld2018}. Furthermore,  delayed communication is connected to higher frustration with and lower liking of chat partners \cite{hwang2019}, mirroring in-person communication, where faster responses are linked to perceived closeness and enjoyment of the conversations \cite{templeton2022}. These findings make communication speed an important yet underappreciated aspect in the quantitative study of reciprocity. In addition to speed, another aspect of the temporal dimension of reciprocity concerns its stability over time. It has been shown that some aspects of social interactions, such as social signature, i.e., how interactions are distributed across contacts, are persistent over time \cite{saramaki2014}. However, it is unclear whether this also applies to reciprocity.

In this paper, we perform, to our knowledge, the first analysis of \gls{RT} balance in \gls{IM} chats. We use \generalvar{n_chats} dyadic chats from the popular platforms WhatsApp and Instagram donated by \generalvar{n_donations} study participants. Data donations were conducted on a privacy-preserving data donation platform \cite{hakobyan2025} that submits only meta data (text length, timestamps and anonymized IDs) without message content. With these data, we address the following three research questions:

\begin{itemize}
    \item[\textbf{RQ1}:] How quickly do people reply in WhatsApp and Instagram? %
    \item[\textbf{RQ2}:] How balanced are the response times between chat partners? %
    \item[\textbf{RQ3}:] Does the response time balance change over time? %
\end{itemize}

We calculated response times and examined their similarity across chat partners in dyadic (two-person) chats. To compare the general response time distributions, we employed a metric based on the Jensen-Shannon divergence (\S \ref{subsec:jsd}). Next, we focused on fast replies and compared chat partners by measuring the difference in their probabilities of responding within five minutes using linear mixed-effect models (\S \ref{subsec:lmm}). Finally, we assessed monthly similarity over time to evaluate the stability of response time similarity.

This paper makes the following key contributions:
    
\begin{enumerate}
    \item We used digital trace data to quantitatively analyze temporal reciprocity in computer-mediated communication.
    \item We collected \generalvar{n_chats} chats across \generalvar{n_donations} donations in a privacy-preserving manner, totaling \datfiltered{n_messages} messages (\S \ref{subsec:data-collection}).
    \item We found that response times are fast across social contexts and platforms: in WhatsApp chats with close contacts and in Instagram chats from a broader social network (\S\ref{subsec:rq1}). 
    \item We showed high response time reciprocity between chat partners: overall distributions aligned closely and the probabilities of fast replies were highly related for chat partners in dyadic chats (\S\ref{subsec:rq2}). 
    \item We found that response time reciprocity persists over time (\S\ref{subsec:rq3}).
\end{enumerate}

\section{Related Work}
\subsection{Measuring Reciprocity with Objective Data}
With the availability of objective trace data from digital communication, several studies have examined reciprocity in more quantitative terms, often using the concept of communication balance as proxy of reciprocity. To this end, several operationalizations have been proposed. 
For instance, \citet{kovanen2010} measured balance in mobile phone networks by looking at how often each conversation partner initiates contact through calls. Extending this idea, \citet{wang2013} compared how likely each party in a dyad is to initiate a call relative to how often they initiate calls with others. \citet{chowdhary2023} defined reciprocal interactions as those where the two parties take turns rather than one person initiating an interaction multiple times in a row. They studied reciprocity in a multichannel scenario using call logs as well as \gls{SMS}, private messages of a university social network, emails, and social media. Finally, instead of contact initiation, \citet{hakobyan2025} focused on interaction volume by measuring word counts of conversation partners in donated WhatsApp and Facebook chats.

Overall, these studies have found that digital communication data is generally characterized by reciprocity yet with a non-negligible fraction of non-reciprocal interactions. For example, \citet{kovanen2010}  found that over 25\% of dyads had one party initiating more than 80\% of calls, while only 14\% of calls were considered non-reciprocal under the operationalization of \citet{wang2013}.   A higher level of call initiation reciprocity was found by \citet{chowdhary2023} for calls (44\%),  online social network messages  (67\%) and \gls{SMS} in (74\%). Instead of analyzing all dyads together, the anonymous authors reported communication balance at the individual user level. They found that the average communication bias was very close to zero for most users, meaning that individuals, on average, sent and received similar number of words.

While previous research has provided valuable insights, it also faces several limitations. In particular, these studies largely focused on forms of communication that do not fully reflect the nature of modern instant messaging. For example, phone calls and SMS were associated with monetary costs, introducing constraints that differed between prepaid and postpaid users. In addition, it is often difficult to determine who initiates contact in instant messaging, as conversations consist of a continuous flow of messages without clearly defined boundaries.
Given the widespread adoption of instant messaging with more than 3.5 billion users worldwide in 2025 \cite{statista-mobile-users}, it is important to account for its specific properties, particularly the speed of communication. As exchanges occur much more rapidly than in earlier communication media, reciprocity in instant messaging is unlikely to be fully captured by aggregate measures such as overall initiation balance or interaction volume alone. Instead, response times may provide a complementary indicator of reciprocity by capturing whether conversation partners respond at similar speeds within a dyad.

\subsection{Temporal dimension of IM communication}
As the name implies, instant messaging is seen as a near-synchronous mode of communication, characterized by rapid exchanges. As early as a decade ago, participants reported replying to chat messages within a few minutes, while others indicated response times of up to an hour \cite{pielot2014}. The quick engagement with instant messages is supported by empirical evidence, with a median message-reading time of 6.15 minutes, although reading may not necessarily imply an immediate response. Studies using data from WhatsApp group chats have confirmed fast communication, with around 60\% of messages receiving responses within one minute \cite{seufert2015, rosenfeld2018}. In work context, people have reported a mean response speed typically not exceeding two minutes \cite{avrahami2008}. %

The temporal aspect of instant messaging is important for two reasons. First, as users often expect fast replies, deviations from this norm can be perceived negatively in the sense of expectancy violations theory \cite{burgoon2015}. This has been shown both in personal \cite{heston2017} and professional \cite{kalman2011} contexts. Violations of expected reply times  may be interpreted as failure to reciprocate communicative effort, thereby linking response timing directly the norm of reciprocity \cite{gouldner1960}. The most extreme case of unreciprocated expectations may be the complete absence of replies, a phenomenon often described as ghosting \cite{buttner2025} and linked to experiences of cyber-ostracism \cite{lutz2023}. 

The second aspect regarding the temporal dimension of reciprocity concerns its dynamics over time. In his seminal work on social network formation, \citet{granovetter1973} used the mutual exchange of favors and help, as a determinant of tie strength, alongside time spent together, emotional intensity, and intimacy. Empirical evidence suggests that reciprocal friendships last more than twice as long on average \cite{hallinan1978}, while, without reciprocation, social ties decay, especially for non-kin relationships \cite{kleinikkink1999}. The persistence of reciprocity remains unexplored in computational social science, although digital traces have been used to examine the dynamics of other aspects of social exchange, such as network characteristics \cite{saramaki2014,roy2022}.

While previous research has examined general response time patterns in 
\gls{IM} \cite{seufert2015,rosenfeld2018} and differences between users and their \textit{aggregated} contacts \cite{hakobyan2025}, response speed has not been examined at the level of dyadic (two-person) interactions. This is a notable omission given that people perceive response speed as an important communicative signal. Interpersonal expectations of reciprocity \cite{burgoon2008} imply that a more focused dyadic analysis  may reveal broadly similar response times within a dyad. The extent to which such patterns emerge in dyadic interactions and remain stable over time is examined in the present work.

\section{Methods}
\label{sec:methods}
\subsection{Data collection \& description}
\label{subsec:data-collection}
\label{subsec:data-description}
Adhering to the ethical guidelines issued by the national psychological associations, chat data were collected using the data donation platform Dona \cite{hakobyan2025}. Using Dona, study participants anonymously donated their chat metadata and received personalized visualizations of their data. Anonymization was achieved user-side by replacing names or phone numbers with \glspl{UUID}. Message contents were aggregated either by word counts or the length of audio messages in seconds before submitting the resulting meta data to the backend, so no identifiable data left participants' devices.
The study received ethics approval from the Bielefeld University ethics committee with application number EUB 2024-098. Our research adhered to the General Data Protection Regulation.

To start the donation procedure, the participants navigated to the Dona website where they received the relevant information about the study and instructions on how to request their data from WhatsApp and Instagram. They agreed to the informed consent and after downloading their data from the platforms, the participants loaded the data into their browsers, reviewed that the data are anonymized and submitted them.
Details on recruitment, the study procedure, and compensation can be found in Appendix \ref{app:study-procedure}.

Depending on the \gls{IM} platform, participants needed to go through very different processes for obtaining their chat data, which also influences the nature of what they donated. WhatsApp chats can be exported in the app immediately, but only one by one and not as a bulk. To reduce the participant burden, participants were asked to only donate the 5-7 chats that they identify as most important to them. At the same time, the data collected from WhatsApp does not contain any information on non-text messages, such as audio, images, or reactions. However, they contain system messages that inform chat participants of changed phone numbers, or encryption, among other events. %

Instagram's data request procedure involved multiple steps and specific user choices that affected the time span and completeness of the donated data. Requested data then became available after a delay of up to one day. In contrast to WhatsApp data, Dona could extract audio messages from the Instagram data. These were anonymized by replacing the actual audio with its length in seconds. Further, Instagram donations also contain timestamps for other media sent, and all chats of a donor. In Instagram's data request procedure, one of the options that participants could choose is the time range of the data download, which defaulted to one year. Therefore, the average Instagram donation time span is around a year, while it is considerably longer for WhatsApp. Due to these differences, we analyzed WhatsApp and Instagram separately. While we compared emerging patterns, differences in results cannot be directly compared and should not be interpreted as platform differences when they could be explained by the differences of the data detailed above.

At the end, the data consisted of the word count/audio message length, and the timestamp in seconds or minutes, depending on platform and operating system. In addition, they have anonymized sender IDs, chat IDs and donation IDs.

During data donation, participants could fill out demographic questionnaires. As no question was mandatory, the number of responses is lower than the number of donations. Participants were mostly young ($Median = 24$ years of age, $SD=6.3$, $\text{range} = 17-57$, $N=93$). 59 participants indicated female gender, 30 male and 3 diverse ($N=92$). Questions on education were not part of all studies and shown only to $N=41$ participants. Of those, 21 finished secondary schooling and 19 obtained some form of higher education.

\subsection{Data processing}
\label{subsec:data-processing}

\begin{figure*}
	\centering
	\includegraphics[width=1.0\linewidth]{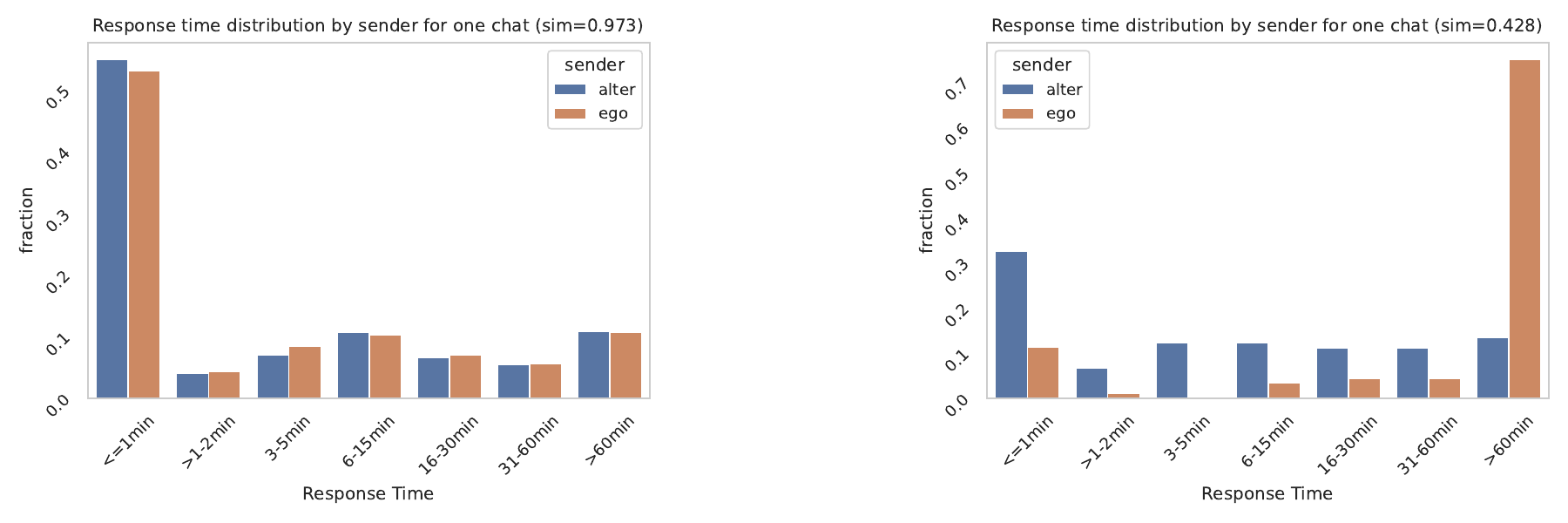}
	\caption{Ego vs. alter \gls{RT} histogram for two specific chats. Left: A chat with high similarity between ego's and alter's response time distribution. Right: A chat with medium similarity between ego and alter, indicating comparatively dissimilar \glsps{RT}.}
	\label{fig:jsd_example}
\end{figure*}

We performed a number of data processing steps encoding assumptions about the data and our research questions for both WhatsApp and Instagram donations, and we argue for them in the following.

Responses are messages relating to previous messages. Without knowing the relations between messages within a chat, we had to assume that every message is a response to the previous message by the other person.
This assumption has less support in group chats where a message can be a response to any other person that previously texted. Hence, we removed chats with more than two participants from our analysis. 

To filter group messages, system messages first had to be excluded from WhatsApp chats as their presence could make a chat appear to have more than two participants (e.g., user A, user B and the system). Removing them ensured that no dyad was mistakenly removed. Therefore, we looked at all triads and identified the sender with the least number of messages. We then decided for a cutoff value of 10 messages and removed all messages of senders with less messages in group chats. A distribution of such message counts for the cutoff justification as well as other cleaning steps that we performed can be found in Appendix \ref{app:preprocessing}. 

\Gls{IM} is used for all sorts of purposes, from selling used goods online to sharing secrets with close friends. To reduce the noise from short-term interactions and focus more on chats with close people, we performed basic filtering steps on the data, removing chats in which a single person wrote less than $10\%$ of the messages and chats that contained less than 100 messages in total. Although of interest, too few group chats remained after filtering to warrant separate analysis, which is why we excluded group chats and relied on dyadic chats only.
We further excluded donations with less than five chats after the other filtering steps.

\subsection{Response time}
\label{subsec:rt}
We defined the \gls{RT} of a message as the time in minutes (rounded up) between the “sent”-timestamps of that message and the previous message. The temporal resolution was rounded to minute-level timestamps used by Android WhatsApp exports for all data sources. An illustration of how \glspl{RT} are created can be found in the Appendix, Figure \ref{fig:rt_def}.

To normalize for some people sending multiple subsequent messages instead of bundling the content into one message, we merged blocks of consecutive messages sent by the same person into one. Omitting such a normalization would have given disproportionate influence to people splitting their messages into multiple smaller ones. For the \gls{RT} of such a block, we used the timestamp of the earliest message of the block.

Meanwhile, for the calculation of the \gls{RT} of the message that replied to this block of messages, the last timestamp of the merged messages is used, so we always used the closest timestamps. In line with this definition, we merged consecutive messages into one after calculating the \glspl{RT}. We also tested aggregation by the latest timestamp and despite numerical differences saw similar response time patterns (Appendix \ref{app:message_aggregation}).
\subsection{Response time similarity}
\label{subsec:jsd}

As a first comparison between \glspl{RT} of egos and alters, we calculated a distribution of each person's \glspl{RT} with seven bins (see Figure \ref{fig:jsd_example}) chosen according to \citet{rosenfeld2018}: 1min, 2min, 3-5min, 6-15min, 16-30min, 31-60min, $>60$min. 
This binning allows handling the otherwise problematic heavy tails of slow responses that are not the focus of this work as well as unequal message counts between chat partners.

We then compared the distributions of ego and alter separately for each chat with a similarity metric based on \glsfirst{JSD}, which itself is a metric based on the Jensen-Shannon divergence. This in turn is a symmetric version of the Kullback-Leibler divergence \cite{lin1991}. 
For the vectors with the fractions of values within the bins for ego and alter, $p$ and $q$, similarity is calculated as 
\begin{equation}
sim(p, q) = 1-\sqrt{\frac{D(p\Vert m)+D(q\Vert m)}{2}}
\label{eq:sim}
\end{equation}
with $D$ being the Kullback-Leibler divergence and $m$ being the pointwise mean of the vectors $p$ and $q$. With our choice of a base 2 logarithm for entropy, this measure is closer to 1 the more $p$ and $q$ are similar and closer to 0 for dissimilar distributions. 

$sim$ does not care about the distance between or ordering of bins, it treats them as categorical. This means that two hypothetical chat partners, one always responding within a minute and the other always taking two minutes, produce a $sim$ value of 0. If one of the chat partners always took 60min, it would also lead to a value of 0. As such, our measure of similarity can be seen as a conservative, upper bound of reciprocity.

Two example distributions of chats, one with high similarity and one with medium similarity, can be seen in Figure \ref{fig:jsd_example}.

\subsection{Five minute RT association}
\label{subsec:lmm}

\begin{figure}[t]
    \centering
    \includegraphics[width=1\linewidth]{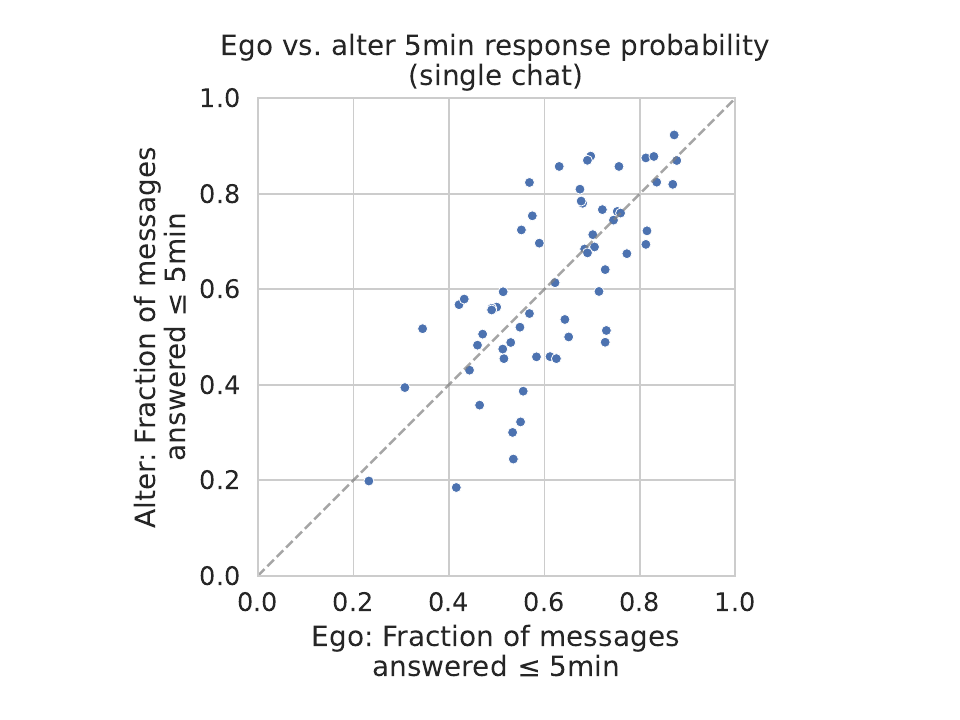}
    \caption{Scatterplot of one ego's five minute response probability vs. alter's five minute response probability for all chats of one donor. Each dot represents one chat of the same ego. Specific ego selected for illustrative purposes.}
    \label{fig:lmm_scatterplot}
\end{figure}

Since responses in \gls{IM} were expected to be fast \cite{pielot2014}, we examined the relationship between ego's and alter's probability to respond within five minutes, in accordance with what is considered a quick response in \citet{rosenfeld2018}. This \glsfirst{RP} was calculated for every chat as the fraction of messages where the \gls{RT} $\leq5$min.%

To examine if there is a linear relationship between the \gls{RP} of egos and their respective alters,  we applied an \gls{LMM}. \Glspl{LMM} model the relationship between some predictors, like the \gls{RT} of one person, and the dependent variable, like the \gls{RT} of their chat partner, by assigning intercepts and slopes to each predictor. Additionally, they can also be used to assign intercepts and slopes to confounding factors that are treated as random effects in this model.
In our case, such a random effect that makes the data non-independent is that multiple chats have the same ego. With one \gls{RP} for ego and one for the alter per chat, we fitted an \gls{LMM} on the egos' five minute \gls{RP}. We used the alters' \gls{RP} as a fixed-effect predictor and accounted for donation-specific variation with random intercepts for each donor. Random slopes for each donor were also tested and discarded due to a lack of improvement. One \gls{LMM} was fitted for WhatsApp donations and one for Instagram. Details for the \gls{LMM} can be found in Appendix \ref{app:lmm}.

As stated in the data processing, to have enough data points for the donor-specific intercepts, we remove donations that have less than five chats after the other processing steps. Due to the differences in WhatsApp and Instagram data, we analyzed and report \gls{LMM} results separated by source.

\pagebreak
\section{Results}
\label{sec:results}
Access to the code used for this paper is available\footnote{https://github.com/mbp-lab/dona-rt2026} and access to the data can be requested by qualified researchers\footnote{https://doi.org/10.5281/zenodo.19369010}.

\subsection{How quickly do people reply in WhatsApp and Instagram chats?}
\label{subsec:rq1}

In total, \datclean{n_donations_ig} participants donated their Instagram data and \datclean{n_donations_wa} donated their WhatsApp data. Although possible, no participant donated data from multiple sources. After the filtering steps described in Section \ref{subsec:data-processing}, \datfiltered
{n_donations_ig} Instagram donations and \datfiltered{n_donations_wa} WhatsApp donations with \generalvar{n_chats_ig} and \generalvar{n_chats_wa} chats remain, respectively, and will be used for our analyses. Together, the two sources provide \datfiltered{n_donations} donations, \datfiltered{n_chats} chats and \datfiltered{n_messages} messages. Detailed statistics of the filtered dataset can be found in Table \ref{tab:general-stats}. A distribution of word counts can be found in Appendix \ref{app:word_counts}.

\begin{table}
	\centering
    \setlength{\tabcolsep}{3pt}
	\begin{tabular}{@{}lllll@{}}\toprule
		& & WhatsApp & Instagram \\
		\midrule
        Donations (n) & & \generalvar{n_donations_wa} & \generalvar{n_donations_ig} \\
        Chats (n) & & \generalvar{n_chats_wa} & \generalvar{n_chats_ig} \\
        \midrule
        Chats & mean & \generalvar{mean_chats_donor_wa} & \generalvar{mean_chats_donor_ig} \\
        per donor & median & \generalvar{median_chats_donor_wa} & \generalvar{median_chats_donor_ig} \\
        & SD & \generalvar{sd_chats_donor_wa} & \generalvar{sd_chats_donor_ig} \\
        & range & \generalvar{min_chats_donor_wa} - \generalvar{max_chats_donor_wa} & \generalvar{min_chats_donor_ig} - \generalvar{max_chats_donor_ig} \\
        \midrule
        Time span & mean & \generalvar{mean_timespan_donor_wa} & \generalvar{mean_timespan_donor_ig} \\
        per donor & median & \generalvar{median_timespan_donor_wa} & \generalvar{median_timespan_donor_ig} \\
        (days) & SD & \generalvar{sd_timespan_donor_wa} & \generalvar{sd_timespan_donor_ig} \\
        & range & \generalvar{min_timespan_donor_wa} - \generalvar{max_timespan_donor_wa} & \generalvar{min_timespan_donor_ig} - \generalvar{max_timespan_donor_ig} \\
        \midrule
        Messages & mean & \generalvar{mean_messages_donor_wa} & \generalvar{mean_messages_donor_ig} \\
        per donor & median & \generalvar{median_messages_donor_wa} & \generalvar{median_messages_donor_ig} \\
        & SD & \generalvar{sd_messages_donor_wa} & \generalvar{sd_messages_donor_ig} \\
        & range & \generalvar{min_messages_donor_wa} - \generalvar{max_messages_donor_wa} & \generalvar{min_messages_donor_ig} - \generalvar{max_messages_donor_ig} \\
        \bottomrule
	\end{tabular}
	\caption{General statistics of people's donated data after filtering but before merging consecutive messages by the same person.}
	\label{tab:general-stats}
\end{table}

\begin{figure*}[t]
	\centering
	\includegraphics[width=1.0\linewidth]{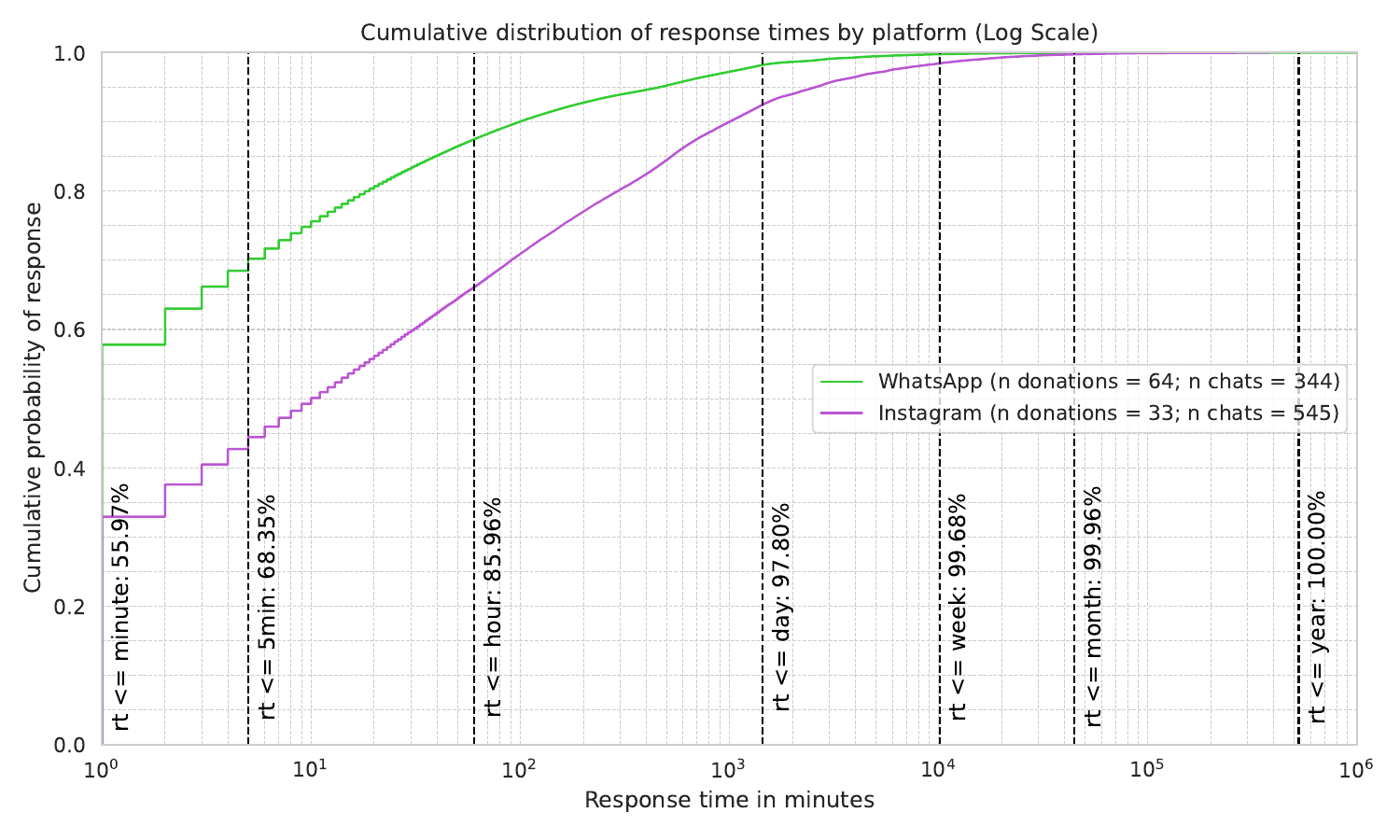}
	\caption{Cumulative distribution of response times over all chats and donors, separated by data source (green for WhatsApp, purple for Instagram). Black dashed lines indicate thresholds of 1min, 5min, 1h, 24h, 1 week, 1 month, 1 year and how many messages have a faster or equal response time (aggregated over both data sources).}
	\label{fig:rt_ecdf}
\end{figure*}

For an overall impression of response speed, we first looked at the cumulative distribution of \glspl{RT}, pooled across all donations of the same platform (WhatsApp and Instagram).
In Figure \ref{fig:rt_ecdf}, it can be seen that people answered 68.88\% of WhatsApp messages and 43.62\% of Instagram messages within five minutes. %
The \gls{RT} distributions also show an extremely heavy tail, with around 0.22\% of WhatsApp and 1.54\% of Instagram messages not having replies within a week.
Overall, we found mostly quick \glspl{RT}.

\subsection{How balanced are the response times of chat partners?}
\label{subsec:rq2}

\begin{figure}
	\centering
	\includegraphics[width=1.\linewidth]{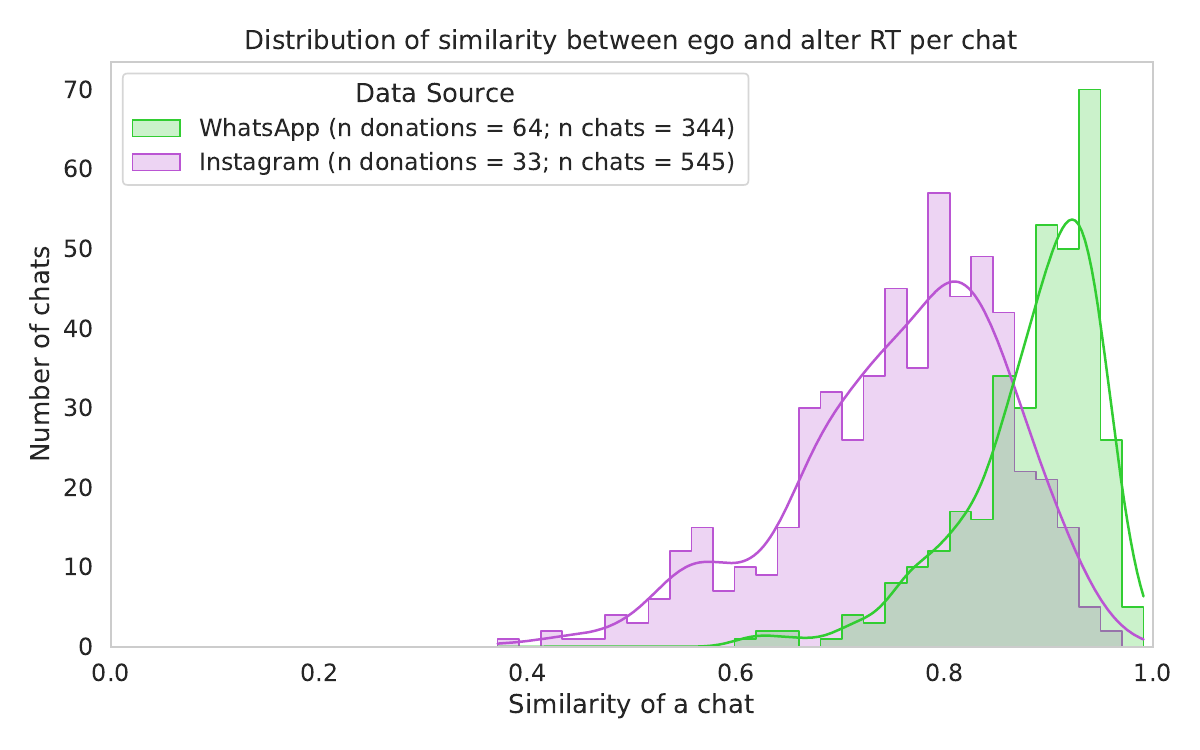}
	\caption{Histogram of per-chat ego-alter similarity. For each of the 889 chats, the similarity is calculated and binned. Higher values indicate more similar response speeds. Separated by data source (green for WhatsApp, purple for Instagram).}
	\label{fig:jsd_hist}
\end{figure}

The balance between chat partners' \glspl{RT} was first examined with our similarity measure. The \glspl{RT} of ego and alter within each chat were binned separately to yield a discrete distribution. We then calculated the similarity between ego's and alter's \gls{RT} distribution using Equation \ref{eq:sim}. Over all chats but separated by source, the similarity values can be found in Figure \ref{fig:jsd_hist}. 
For WhatsApp, per-chat \gls{RT} distributions  between egos and alters are close to similar with $Mean=\jsdvar{mean_wa}$, $SD=\jsdvar{sd_wa}$. Instagram chats have a mean similarity of $\jsdvar{mean_ig}$ and a standard deviation of $\jsdvar{sd_ig}$. For both Instagram and WhatsApp, %
 \gls{RT} similarity values fall almost entirely into the higher half of the similarity range (0 to 1), indicating largely aligned distributions between chat partners. 

For a more focused analysis considering the fast response speed, we employ an \gls{LMM}, allowing us to examine if alters' probability of a fast reply (\gls{RT}$\leq5$min) is associated with the fast \gls{RP} of their corresponding ego. Since the majority of messages were answered quickly, and the distribution comparison showed similarity between chat partners, we also expected an association in the fast \gls{RP}. An example of the five minute \gls{RP} data from one ego's chats can be seen in Figure \ref{fig:lmm_scatterplot}.

The model results in Table \ref{tab:lmm} show that for WhatsApp, the slope is 0.787. This means that on average, an increase of 10 percentage points in the alters' probability of responding within five minutes is associated with an 7.87 points increase in egos' fast reply probability. The slope indicates that chat partners have comparable tendencies to respond quickly.
Despite Instagram chats appearing less balanced in terms of \gls{RT} similarity, the association is even higher for Instagram than for WhatsApp, showing a slope of 0.797. 
The association of \glspl{RT} between egos and their alters is significant for both the WhatsApp and the Instagram model.

Both the similarity and the \gls{LMM} analysis showed comparable response times between chat partners with only small platform-specific differences. %

\begin{table*}
    \centering
    \begin{tabular}{@{}llllllll@{}}\toprule
     & Predictor & Coef. $\beta$ & SE($\beta$) & $95\%$ CI & DF & $t$ & $p$ \\
     \midrule
    WhatsApp & (Intercept) & \walmm{intercept_est} & \walmm{intercept_se} & [\walmm{intercept_ci_lo}, \walmm{intercept_ci_hi}] & \walmm{intercept_df} & \walmm{intercept_t} & $<$ 0.001 \\
    ($N=344$) & Alter's response time & \walmm{rt_prev_alter_est} & \walmm{rt_prev_alter_se} & [\walmm{rt_prev_alter_ci_lo}, \walmm{rt_prev_alter_ci_hi}] & \walmm{rt_prev_alter_df} & \walmm{rt_prev_alter_t} & $<$ 0.001 \\
    \midrule
    Instagram & (Intercept) & \iglmm{intercept_est} & \iglmm{intercept_se} & [\iglmm{intercept_ci_lo}, \iglmm{intercept_ci_hi}] & \iglmm{intercept_df} & \iglmm{intercept_t} & $<$ 0.001 \\
    ($N=545$) & Alter's response time & \iglmm{rt_prev_alter_est} & \iglmm{rt_prev_alter_se} & [\iglmm{rt_prev_alter_ci_lo}, \iglmm{rt_prev_alter_ci_hi}] & \iglmm{rt_prev_alter_df} & \iglmm{rt_prev_alter_t} & $<$ 0.001 \\
    \bottomrule
    \end{tabular}
    \caption{Linear mixed-effects model for the probability of ego's response within one minute. No transformation or standardization.}
    \label{tab:lmm}
\end{table*}

\subsection{Does the response time balance change over time?}
\label{subsec:rq3}
\begin{figure}
	\centering
	\includegraphics[width=1.0\linewidth]{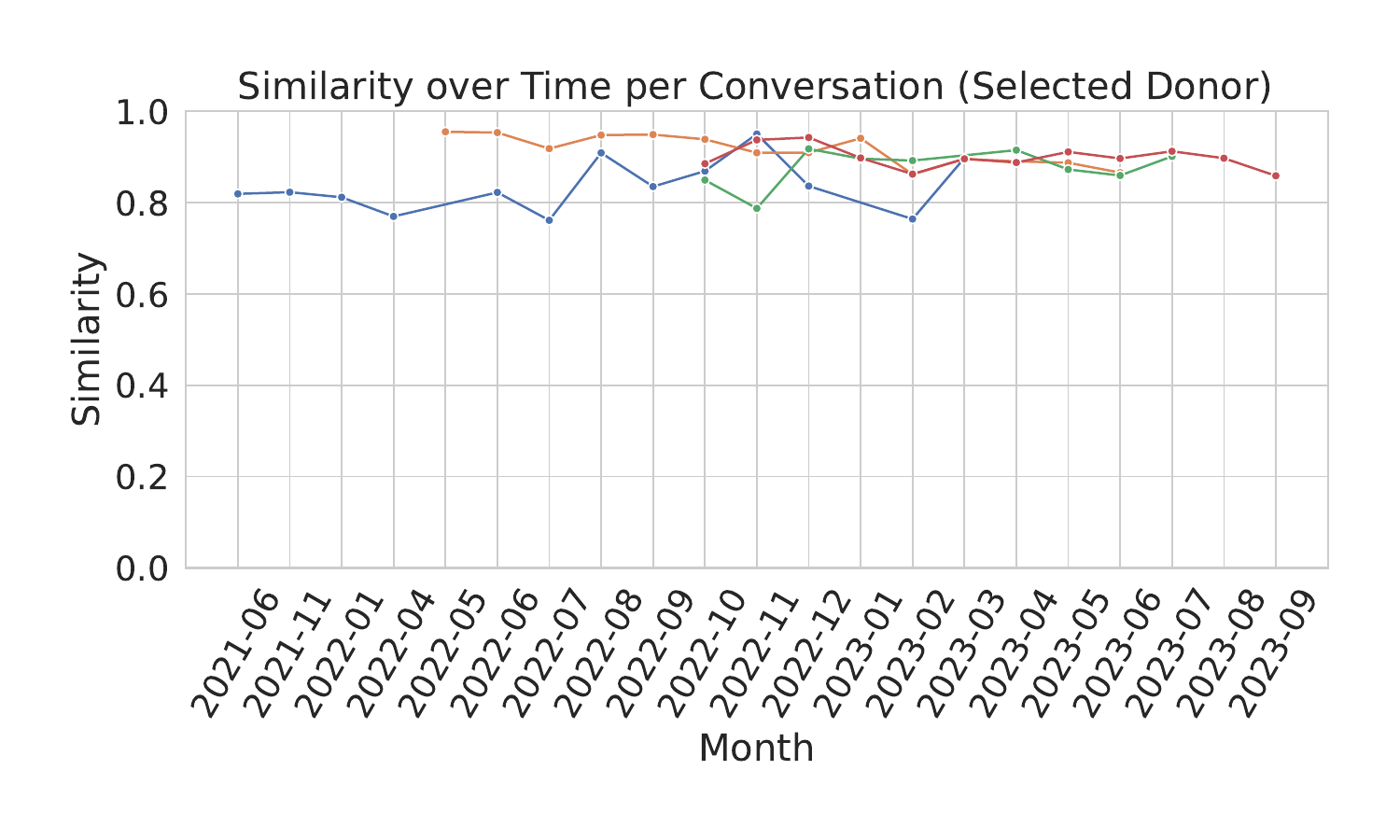}
	\caption{Similarity for chats of one Instagram donor over time. Similarity was calculated for each month with more than 70 messages and only chats with 5 or more months of data are shown. Specific donor selected for illustrative purposes.}
	\label{fig:jsd_temporal_example}
\end{figure}

While we showed that the response times of people are generally rather similar, the previous analysis did not capture changes over time. Examining temporal variations can reveal whether this similarity is persistent or only balanced if aggregated over longer time spans.

First, we calculated the similarity separately for each month and each chat. This required a lower bound on the number of messages sent by each person within that month and chat for acceptable sampling error. We empirically evaluated different threshold choices and decided to only include months with at least 70 messages by each chat partner as a tradeoff between noisy estimates and excluding too many data points. The empirical derivation can be found in Appendix \ref{app:JSD_temporal}.

For a single donor, the per-chat similarity over time can be seen in Figure \ref{fig:jsd_temporal_example}. Values stay above 0.5 throughout the whole donated period, but fluctuate in the range of 0.5 to 1.0. To check if this trend of variations extends across chats, the dispersion of similarity values of each chat was quantified with the \gls{MAD}. It describes the median of each data point's deviation from the chat's median similarity and is employed because of its robustness with few data points and outliers that might be present due to the noise introduced with some months having only few more than 70 messages.

\begin{figure}
	\centering
    \includegraphics[width=1.0\linewidth]{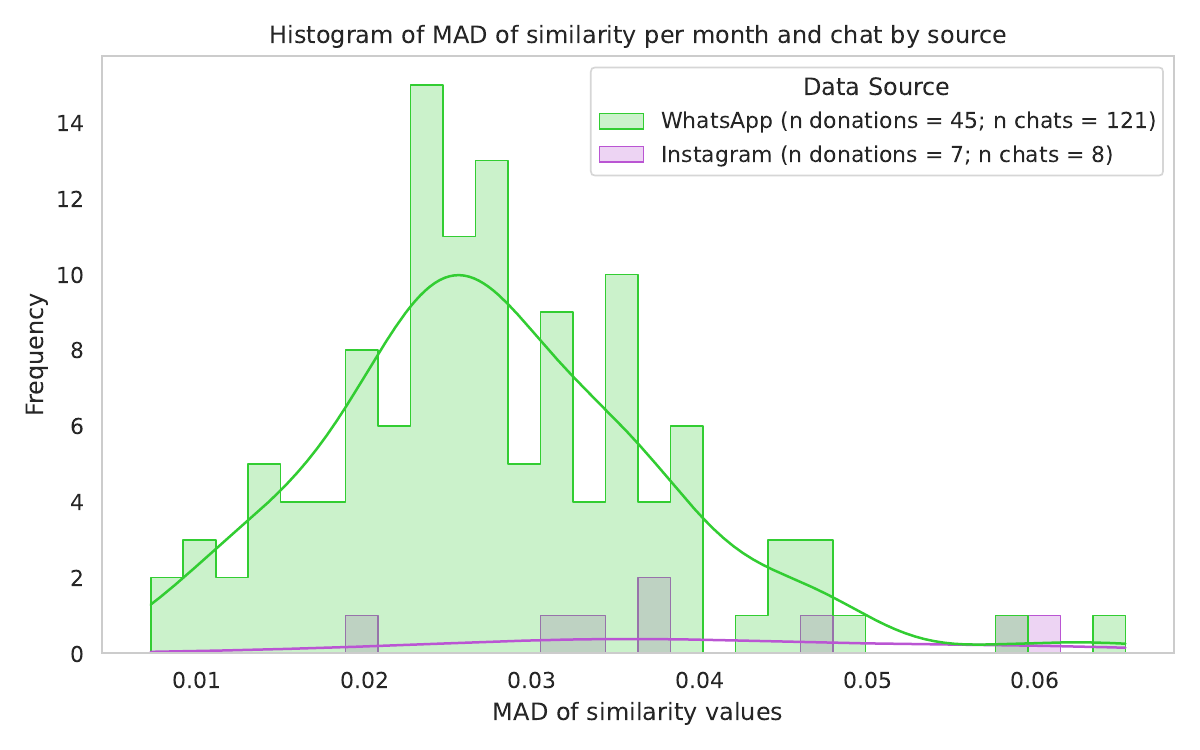}
    \caption{\gls{MAD} of each chat's similarity over time. For each chat with five or more months of messages and at least 70 messages per month, similarity was calculated per month. The dispersion of each chat's similarity was calculated using the \gls{MAD} and binned separately for WhatsApp and Instagram.}
	\label{fig:rt_mad}
\end{figure}

These \gls{MAD} values are plotted in a histogram in Figure \ref{fig:rt_mad}. Every chat's \gls{MAD} falls into the range of 0.01 to 0.07. Compared to the per-month similarity values, which range from 0.35 to 1.0, the lower \gls{MAD} values show small variation. This suggests that, apart from a few outliers, the \gls{RT} balance is stable over time.

\section{Discussion}
\label{sec:discussion}
In the analysis of response times in instant messaging, we found that chat partners typically reply to each other quickly, with most messages being responded to within five minutes. %
These findings confirm expectations from the literature of mostly fast response times in instant messaging \cite{seufert2015,rosenfeld2018,pielot2014}. Our findings revealed that chat partners exhibit similarity both in terms of general response times and when the analysis was restricted to fast responses. Interpreting this dyadic similarity as a proxy for reciprocity, these results are consistent with previous studies that found largely reciprocal communication in digital trace data \cite{kovanen2010, wang2013,chowdhary2023}.     %

In addition to aligning with prior work, our approach introduces a measure of temporal reciprocity that may be better suited to instant messaging communication. Using response speed, it brings a key aspect of messaging into focus that is also subject to interpersonal expectations. Just as reciprocity is a fundamental social norm \cite{gouldner1960}, expectations about appropriate response times have been described as "chronemic social norms" \cite{kalman2021}, with long response delays perceived negatively in the sense of expectancy violation theory \cite{burgoon2015}. Beyond response speed, there are also other aspects that shape messaging impressions more generally, such as text length, as shown by associations between the use of abbreviations and perceptions of insincerity \cite{fang2025}. Examining such effects, however, typically requires access to message content, which conflicts with the strong privacy-preserving focus of this work. Some prior work has indeed used message length from de-identified data as a proxy for interaction balance (anonymous authors). However, there is no evidence that text length elicits strong social expectations comparable to those associated with response speed.  Mismatch expectations in response speed is known to cause communication friction \cite{chou2022}, reduce interpersonal attraction \cite{heston2017}, credibility \cite{kalman2011} and lead to feelings of cyber-ostracism \cite{lutz2023}.

We also investigated the temporal aspect of reciprocity in terms of its stability over time. We found that it has little monthly variation. Given the role of reciprocity as a fundamental interaction feature, this results seems rather unsurprising. However, it is worth noting that reciprocity may fluctuate at finer temporal scales that are not captured by our monthly aggregation. A study of e-mails in a work environment showed that people adapt their response times to match the response times of others \citep{tyler2003}. In an explicitly arranged task where one participant shared information with another via text, a deliberate imbalance of word counts and message length was created, yet participants adapted their messaging pattern to restore balance \cite{guydish2022}.  Reciprocity may appear balanced when averaged over longer periods, such as months, but become more asymmetric at shorter time scales, for example, during periods of stress from one interaction partner. Fluctuations may also occur at much longer time scales than those covered by our data. It has been shown that reciprocity within family relationships shifts over the life course, with children receiving more support early in life and parents receiving more as they age \cite{kleinikkink1999}.  Such dynamics were beyond the scope of the present analysis, which focuses on establishing a temporal marker of reciprocity in the first place. Future work could build on this foundation by examining finer and longer-term temporal variations. This would allow to investigate changes in reciprocity over time enabling longitudinal, data-driven analyses of how changes in reciprocity relate to aspects of relationship, such as closeness and persistence.

We found comparable results across platforms and social contexts. %
WhatsApp donations only contained 5-7 chats as the donor was explicitly instructed to select the most important chats. This was done to reduce participant burden as WhatsApp export function requires chats to be exported individually. Due to this selection process, WhatsApp chats can be assumed to capture the core social network of the participants \cite{dunbar2015}. By contrast, Instagram included data allowed the export of the entire online network, capturing also social ties beyond the core network. The data in our analysis were filtered for quality, e.g. by ensuring that only chats containing at least 100 messages  were included. This means that temporal reciprocity manifests in sustained interactions and does not seem to be affected by idiosyncrasies of specific messaging platforms. This observation aligns with prior work showing that fundamental social interaction patterns are reflected in digital trace data \cite{dunbar2015, maccarron2016}. It remains to be seen whether more fine-grained temporal analysis will reveal higher reciprocity for closer ties, as shown in the previous work based on call initiation \cite{kovanen2010} and higher contact frequencies \cite{sutcliffe2012}. In this regard, a limitation of the present study is that we did not explicitly assess the perceived reciprocity or closeness for individual chats of the donor , although our filtering steps tended to retain chats between closer contacts.

Although temporal reciprocity appears to generalize across the examined platforms, concerns remain regarding generalization beyond the study sample. Our sample consisted of 97 participants with a median age of 24 years. The mostly young age may have resulted in higher typing and reading speed compared to a sample encompassing a wider age range. For example, older people were shown to type slower when using smartphone keyboards \cite{smith2015}. In addition, different age groups may be affected differently by habits, working or study hours.  As we do not have demographic information on the contacts of the study participant, further analysis regarding age and other demographic characteristics was not feasible despite their potential relevance. The sample size was influenced by the challenges inherent in data donation studies \cite{pfiffner2024} and by resource constraints \cite{lakens2022}. Future work should examine the generalizability of these findings not only in larger samples but also in samples that are more balanced with respect to age.

While temporal reciprocity has the advantage of being grounded in social behavior, dealing with response times is not without its practical challenges. We found that the response time distributions are heavy-tailed. This stems from burstiness, referring to the tendency for human activity in general \cite{goh2006} and digital communication in particular \cite{krings2012a} to occur in irregular, clustered bursts. In the response time data, burstiness was evident across different temporal scales, with inter-event times exhibiting heavy-tailed distributions from short to long intervals.  %
Discretized response time analysis through the binning intervals introduced by \citet{rosenfeld2018} helped mitigate the effects of burstiness while enabling analysis on behaviorally meaningful time scales of social interaction. Given that most responses occurred quickly, focusing on the probability of responding within five minutes appeared to be a sensible choice.  %

In this work, we chose to limit the scope to response time analysis without considering its relationship with text length.  A potential concern is that message length may automatically affect response times because reading and typing take time. While true in principle, we expect these execution times to be negligible at the behavioral time scale considered here. Average mobile typing speed is approximately 36 words per minute \cite{palin2019} and reading is generally faster \cite{brysbaert2019}. In our sample, less than $5\%$ of messages contained more than 20 words (see Appendix Figure \ref{fig:wc_hist}), so the time required for typing was likely shorter than the resolution of our data (minute).  We consider cognitive (e.g., deliberation about what to reply) and situational factors (e.g., availability) to play a larger role than the time required to read or type longer messages. Although this cannot be investigated using the privacy-preserving data of our sample, this  points to an important direction for further investigation.

We showed that instant messaging communication is characterized by fast, reciprocal exchanges that exhibit a stable level of reciprocity over time. Our approach provides a quantitative metric that can support efforts in computational social science to better capture the dynamics of social interaction in computer-mediated communication. Future work could use the temporal reciprocity proposed here to examine associations between reciprocity and relationship characteristics, such as its strength. Moreover, this method could be compared to other quantitative measures of reciprocity in dyadic chats (e.g. text balance) as well as subjective perceptions. This would further validate the approach beyond existing literature that establishes the relevance of response timing.

\section{Conclusion}
In this paper, we introduced response time as a quantitative marker of reciprocity in fast-paced instant messaging communication. Analyzing 3.4 million anonymous WhatsApp and Instagram messages we showed that dyadic chats are characterized by rapid replies that are strongly aligned between chat partners in terms of response time similarity. Moreover, we observed a persistence of this alignment over time suggesting that response time balance reflects a stable interactional norm within dyads. By combining a socially meaningful aspect of instant messaging with digital trace data, our approach enables longitudinal analysis of reciprocity and how it relates to interpersonal ties, such as their strength and persistence.

\bibliographystyle{aaai2026}
\bibliography{bibliography}
\appendix
\section{Preprocessing}
\label{app:preprocessing}
There are two groups of preprocessing steps that we conduct. The first is data cleaning, which is concerned with removing erroneous data. The second is transforming and filtering, which only retains data relevant for our investigations and transforms it into a suitable format. This latter step is described in Section \ref{sec:methods}.

\begin{figure}
    \centering
    \includegraphics[width=1.0\linewidth]{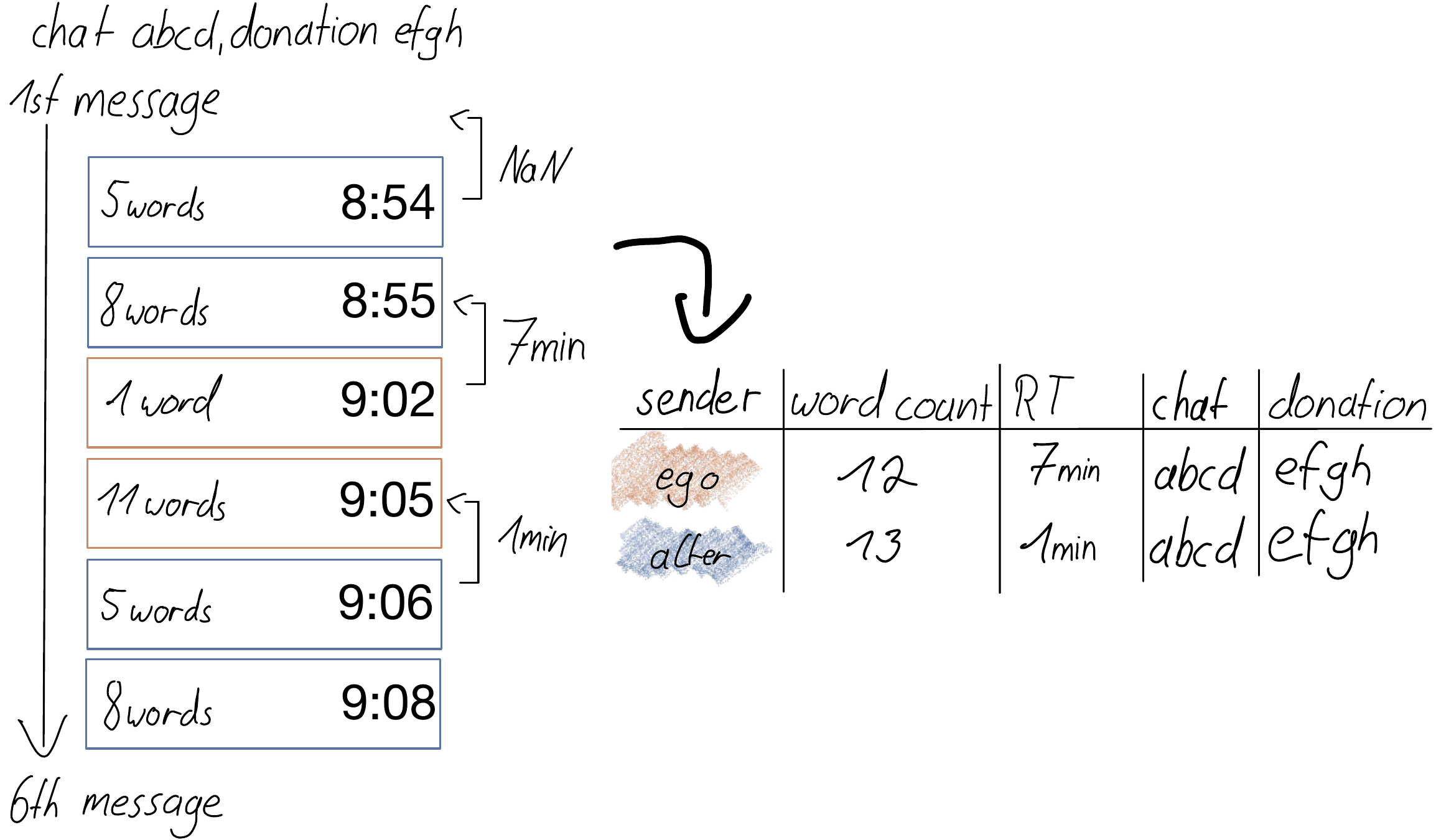}%
    \caption{Schematic depiction of \gls{RT} generation from a chat. For every series of consecutive messages from ego (orange) or alter (blue), one data point is created, using the summary of words and the shortest response time from all messages of the series. The first message in a chat does not create a data point.}
    \label{fig:rt_def}
\end{figure}

\subsection{Data Cleaning}
\label{app:data-cleaning}

\begin{figure}
    \centering
    \includegraphics[width=1\linewidth]{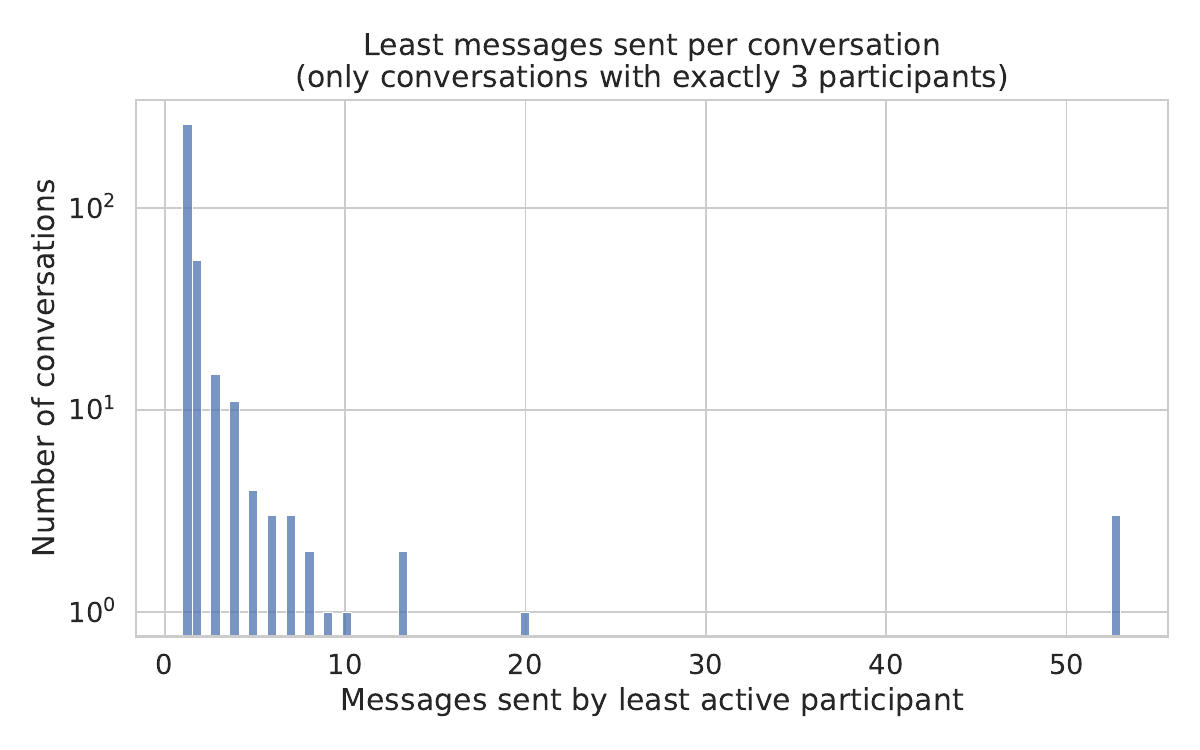}
    \caption{Histogram of lowest message count over all senders for each triad.}
    \label{fig:system_message_hist}
\end{figure}
The steps performed to improve data quality start with removing test donations. These have a known fingerprint, consisting of the number of days encompassed in the donation and the number of messages. Based on this known fingerprint of our test files, we remove all such donations.

In a similar manner, we also use a donation's fingerprint values of earliest and latest datetime as well as the message count to remove duplicate donations. Note that this is a rather exact match and does not protect against deliberate manipulation.

For reasons unknown to us, 67 messages had dates after the date of donation. Those messages were assumed to be erroneous and also excluded.

In WhatsApp chats, some messages are automatically created system messages. These inform about events such as encryption or phone number changes, and are not written by a chat partner. Filtering out system messages exactly requires knowledge of the content, which is language-dependent \cite{kohne2023}. As we do not collect this information, our approach is to filter out the person with the smallest contribution to a chat of more than two people if their contribution is lower than a certain threshold. For all triads, the largest number of messages contributed by the person with the least number of messages in a chat was 53 messages. Based on a histogram of the number of messages sent by the least active participant in a triad (see Figure \ref{fig:system_message_hist}), we chose 10 messages as the cutoff.

In some Instagram donations, the wrong person was identified as the donor. This was corrected by checking which senders were present in the highest number of chats. If there was a sender that was present in more chats than the current candidate, this sender was deemed the new donor. In total, 20 donations were updated this way.

\section{Parameters for \glsentrylong{JSD} over time}
\label{app:JSD_temporal}
\begin{figure}
	\centering
	\includegraphics[width=1.0\linewidth]{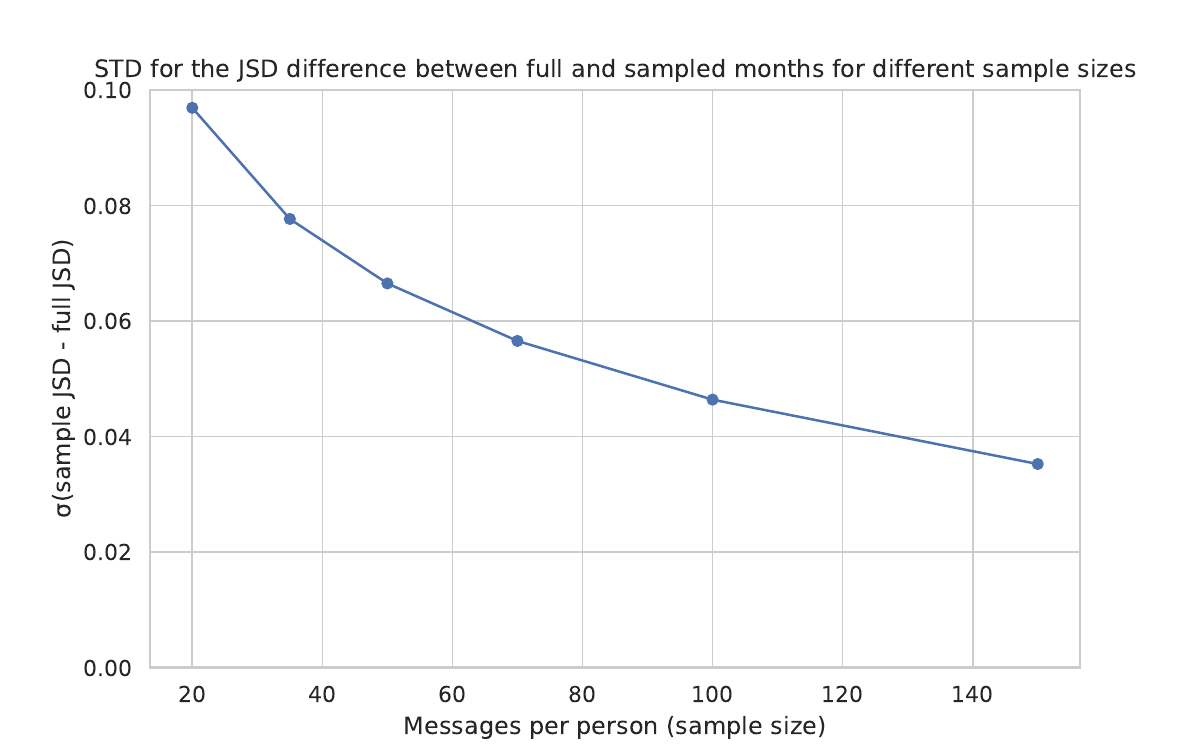}
	\caption{Standard deviation of the differences of \gls{JSD} calculated on subsampled messages vs. on all available messages for all months-chat combinations with more than 200 messages.}
	\label{fig:JSD_std_sampling}
\end{figure}

The minimum number of messages each person sent within a month so that \gls{JSD} could be calculated with acceptable noise was determined empirically. We first selected all months across all conversations with 200 or more messages per person. Then, we calculated the JSD between the response time binning of ego and alter for those months. We continued by sampling a number of messages from each person for each month and chat. The sampling numbers tested were partly informed by heuristics such as $k=\sqrt{n}$, $5k$, and $10k$ (with $k=7$ being the number of \gls{RT} bins) and ranged from 20 to 150. For the \gls{JSD} calculated on these samples, we computed the difference to that of the full number of messages. We repeated this process for a total of 400 times for each month, chat, and sampling number before finally computing the \gls{SD} over all the \gls{JSD} differences. The results can be seen in Figure \ref{fig:JSD_std_sampling}.
Based on the \gls{SD} values obtained, we decided for 70 messages as this would allow us to label differences of around 0.15 as above a 95\% noise threshold while still keeping an acceptable amount of chats (going from 889 to 180).

\section{Participant recruitment}
\label{app:study-procedure}
The data used in the present paper was pooled from multiple studies. In those studies, participants were asked to fill out questionnaires in addition to the data donation, which differed based on the study goals.

\begin{figure*}
    \centering
    \includegraphics[width=0.75\linewidth]{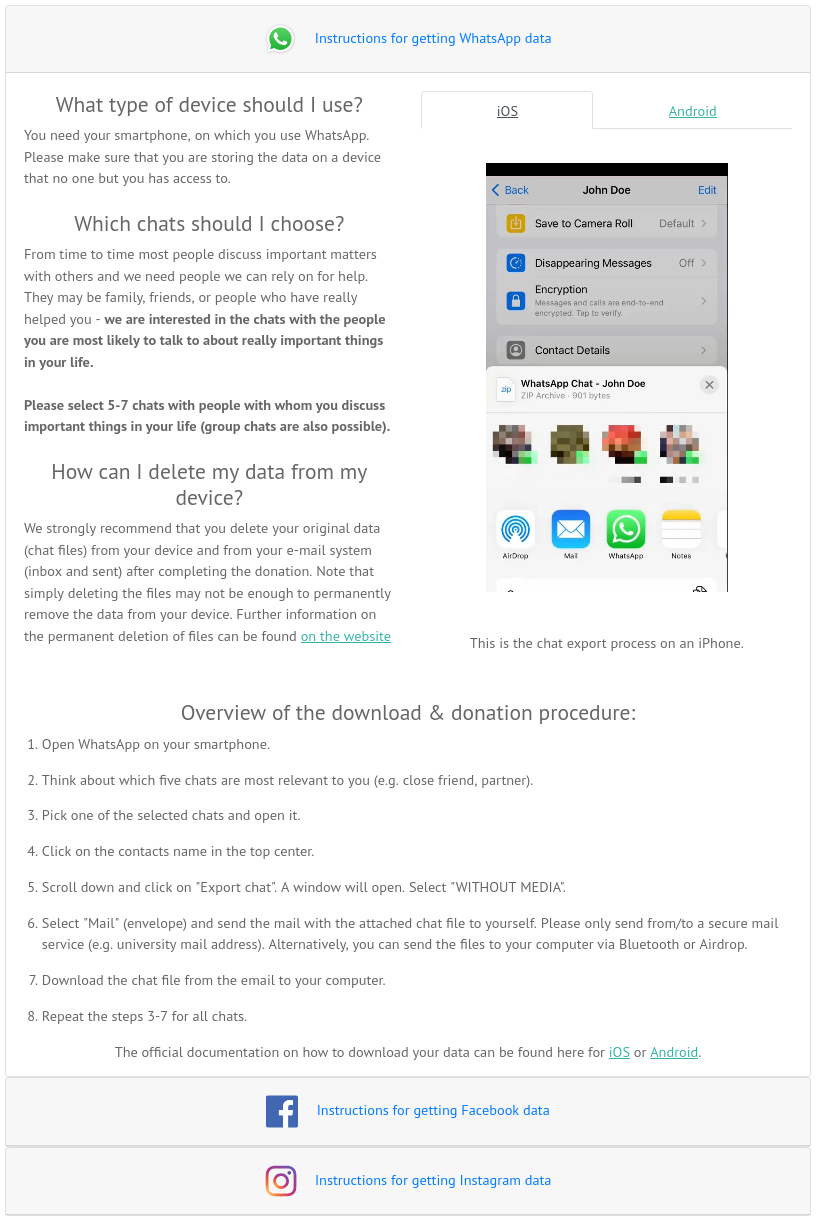}
    \caption{Participant instructions for obtaining their WhatsApp data.}
    \label{fig:wa-instructions}
\end{figure*}

To donate data, interested participants that navigated to the website of Dona were greeted with information about why their participation matters, what they gain from doing so, and how the data donation and anonymization process works. After agreeing to the informed consent, participants could read about how they can get their chat data from the supported platforms. Participants then obtained their WhatsApp or Instagram data from the respective platforms, which was supported by specific instructions on the Dona website, and loaded it into their web browsers, where it was anonymized locally. Because WhatsApp chats have to be downloaded individually, participants were only asked for their 5-7 most important chats. The instructions that participants get for obtaining WhatsApp data can be found in Figure \ref{fig:wa-instructions}.

After participants reviewed their data to be completely anonymized, they could choose to send the data to the Dona backend. There, it was stored and used to generate personalized, interactive infographics on aspects such as response times, chat balance, and chat intensity. Participants could examine these feedback plots and download them. Finally, they could sign up for compensation.

Participants took between 15 and 60 minutes for the studies and were compensated with x anonymous currency. The time required for the data donation part was not tracked separately.

People became aware of the study through flyers, Instagram ads, in-person events, and word-of-mouth recommendations.

\section{Linear mixed-effects model details}
\label{app:lmm}

In accordance with the best practices for \glspl{LMM} recommended by \citet{meteyard2020a}, we build our model from simple to complex, meaning that we start with a simple model and add more factors if they improve the model. Starting from the formula \texttt{rt\_ego$\sim$ rt\_alter+(1|donor\_id)}, we tested random slopes per donor, time of day and day of week, but found no significant improvement and omitted them from the final model.

\begin{figure*}%
    \centering
    \begin{subfigure}[t]{0.48\textwidth}
        \centering
        \includegraphics[width=\linewidth]{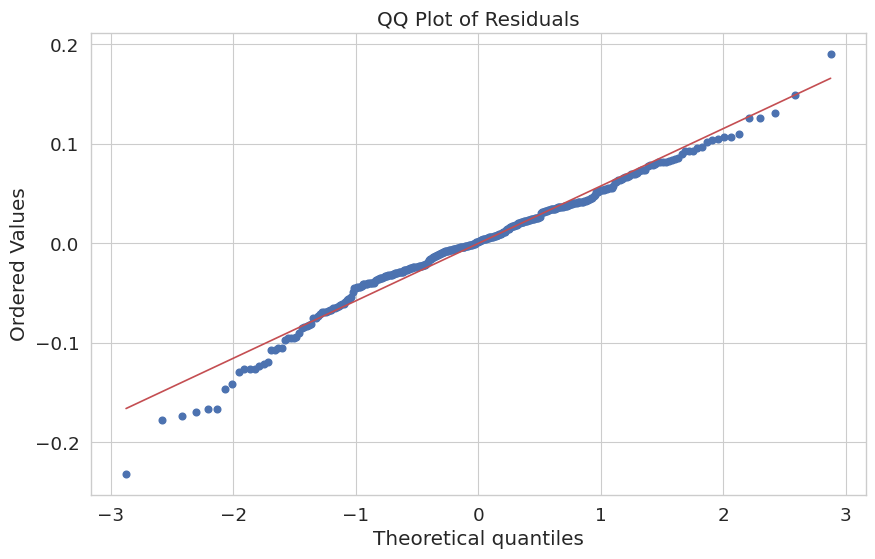}
        \caption{WhatsApp}
        \label{fig:wa_residuals_qq}
    \end{subfigure}%
    \hfill
    \begin{subfigure}[t]{0.48\textwidth}
        \centering
        \includegraphics[width=\linewidth]{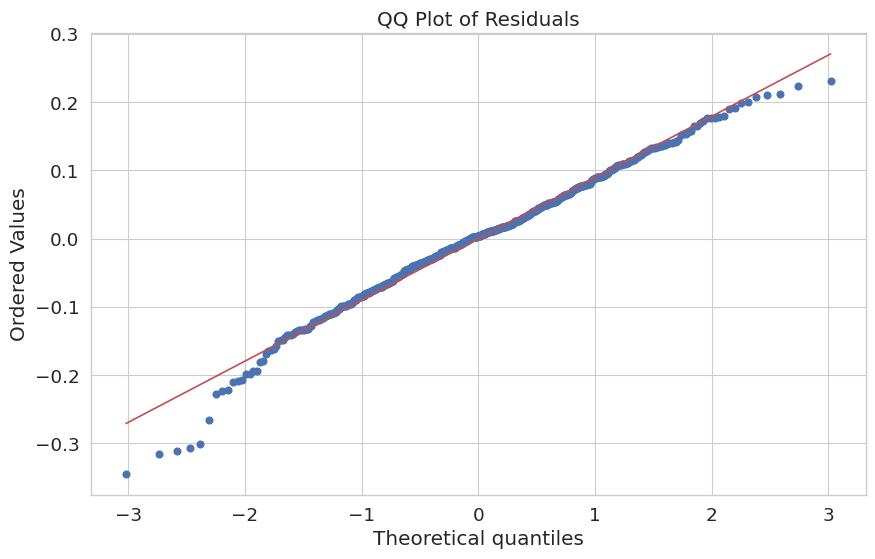}
        \caption{Instagram}
        \label{fig:ig_residuals_qq}
    \end{subfigure}%
    \caption{Q-Q plot of residuals of the \glspl{LMM} fit on egos' probability of responding within 5 minutes based on the same probability of their alters.}
    \label{fig:qq_plots}
\end{figure*}

All data processing steps are described in Section \ref{sec:methods}. No transformations have been performed as the data already met assumptions of normality. Residual Q-Q plots can be found in Figure \ref{fig:qq_plots}.

Packages and versions for modeling can be found in the code provided in Section \ref{sec:results}.

\section{Word Count Distribution}
\label{app:word_counts}
\begin{figure}
	\centering
	\includegraphics[width=1.0\linewidth]{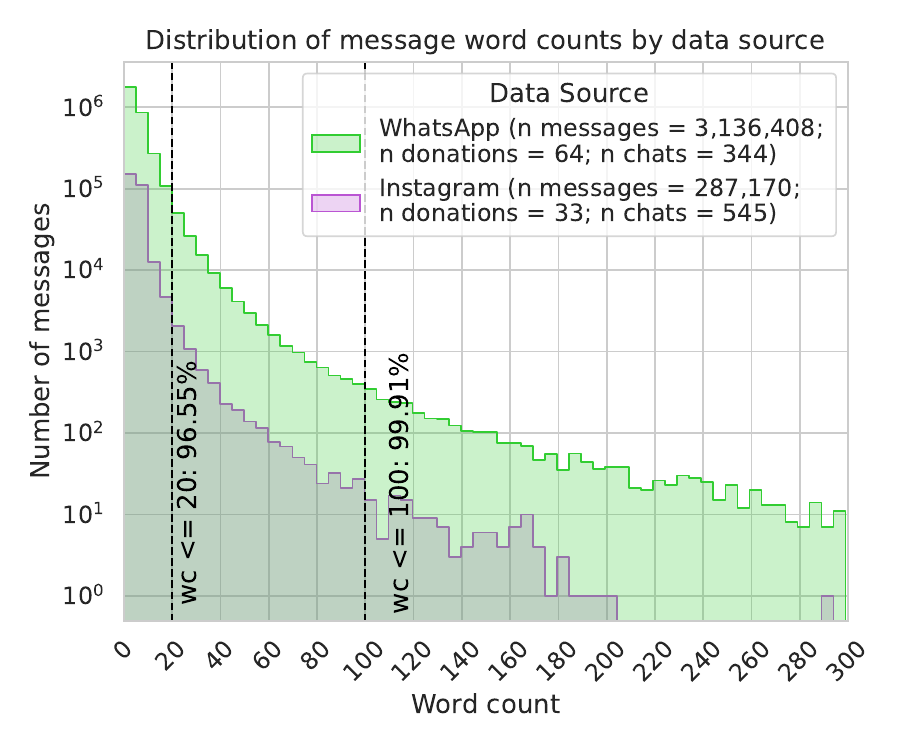}
	\caption{Logarithmic histogram of word counts per message, separated by platform. 272 messages contained more than 300 words and were omitted for the histogram but retained for the calculation of percentages. No merging of consecutive messages by the same sender was performed.}
	\label{fig:wc_hist}
\end{figure}

A histogram of word counts without merging consecutive messages by the same sender can be found in Figure \ref{fig:wc_hist}. Less than $4\%$ of messages contained more than 20 words.

\section{Influence of Message Aggregation}
\label{app:message_aggregation}
Consecutive messages by the same sender are aggregated into blocks to account for texting behavior where every message contains one thought and multiple messages are sent consecutively. Omitting aggregation would have given disproportionate influence to people splitting their messages into multiple smaller ones.

\begin{figure*}
    \centering
    \includegraphics[width=0.8\linewidth]{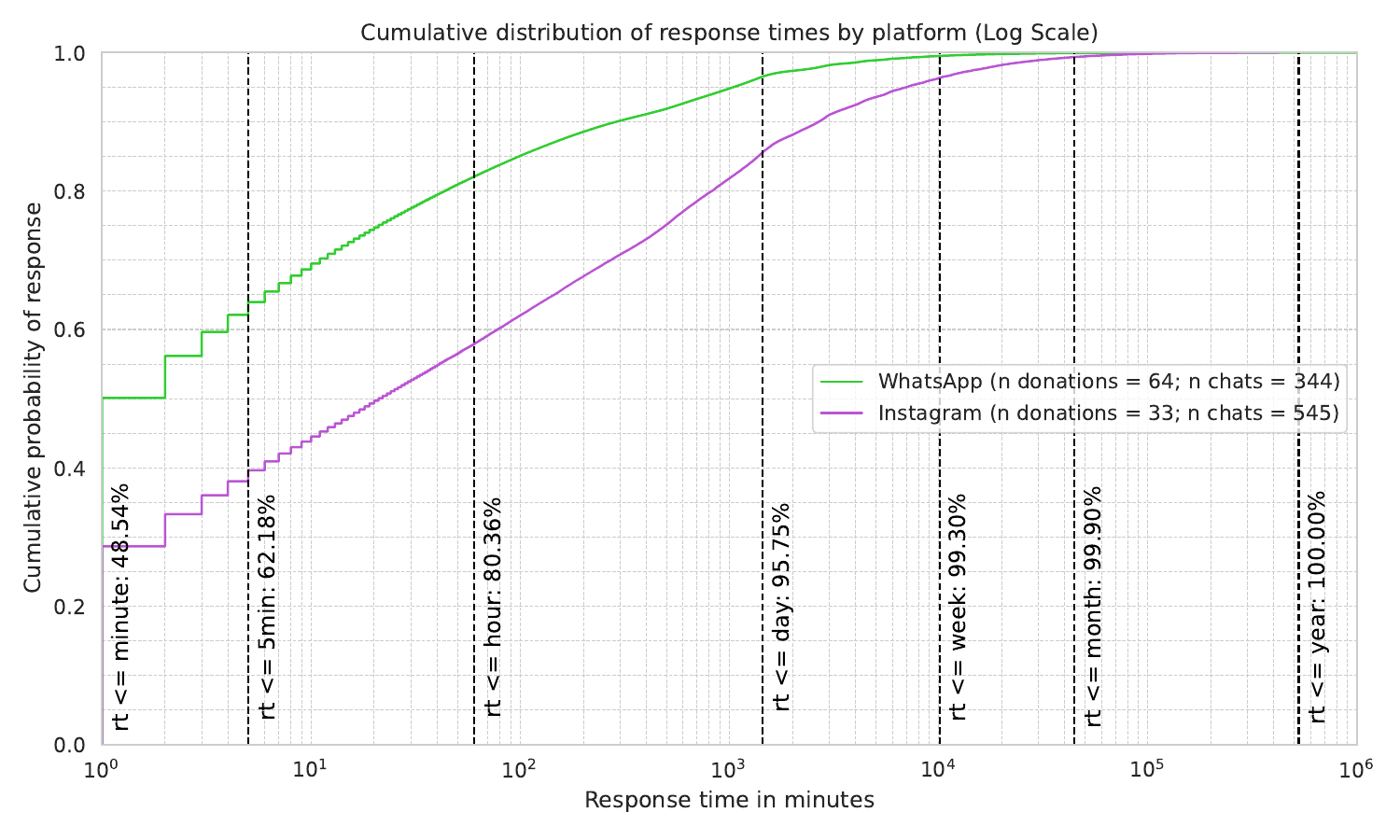}
    \caption{Cumulative distribution of response times over all chats and donors, separated by data source (green for WhatsApp, purple for Instagram). Black dashed lines indicate thresholds of 1min, 5min, 1h, 24h, 1 week, 1 month, 1 year and how many messages have a faster or equal response time (aggregated over both data sources). Compared to Figure \ref{fig:rt_ecdf}, blocks of consecutive messages by the same person are aggregated by the last instead of the first timestamp within this block.}
    \label{fig:rt_ecdf_keeplast}
\end{figure*}

When messages were aggregated, the earliest timestamp was used to calculate the gap to the previous block of messages, resulting in the shortest \gls{RT} of the whole block. This interpreted the first message as the reply, while later messages might be the start of a new conversation. As our analysis focused specifically on replies, new conversations are of no interest and were discarded -- in the case of aggregation implicitly. If we would assume that the last message of a block is the actual response and there are no new conversations, we could use the last timestamp instead. A cumulative distribution similar to Figure \ref{fig:rt_ecdf} but with blocks aggregated with the last timestamp can be found in Figure \ref{fig:rt_ecdf_keeplast}.

\end{document}